\documentstyle[floats,twocolumn,prl,aps]{revtex}

\input epsf
\input rotate

\begin{document}

\draft

\title{Numerical Simulation of Three-Dimensional Dendritic Growth}

\vskip 0.5 cm

\author{Alain Karma and Wouter-Jan Rappel}

\address{Department of Physics and Center for Interdisciplinary
Research on Complex Systems,\\ Northeastern University, Boston,
Massachusetts 02115}

\author{\parbox{397pt}{\vglue 0.3cm \small
Dendritic crystal growth in
a pure undercooled melt is simulated quantitatively
in three dimensions using a phase-field approach.
The full non-axisymmetric morphology 
of the steady-state dendrite tip 
and $\sigma^*$ are determined 
as a function of anisotropy for a crystal with a cubic symmetry. 
Results are compared to experiment and
used to critically test solvability theory. (June, 1996).
}}
\maketitle

Dendritic growth has been a central problem
in pattern formation \cite{LanHou}
and metallurgy \cite{KurFis} for many years.
Considerable theoretical progress has been achieved
which has led to the development of solvability
theory to determine the steady-state operating state
 of the dendrite tip
(i.e. tip velocity $V$ and tip radius $\rho$) 
\cite{LanHou,Kesetal:geom,Benms,Lan:sol,KesLev,BarLan,BenBre}.

Paradoxically, our understanding of pattern selection
in three dimensions (3D) has remained uncertain,
especially with regards to experiment. This is 
due to the fact that it has so 
far remained too difficult to simulate
reliably the equations of dendritic growth in 3D
even on current supercomputers. 
This, in turn, has prevented:
(i) to test whether the global attractor of the growth dynamics is
indeed the steady-state needle crystal predicted 
by solvability theory, and (ii) to test the predictions
of this theory which are themselves only approximate 
in 3D \cite{KesLev,BarLan,BenBre}.
Consequently, it has remained unclear whether 
existing disagreements between solvability theory and experiment 
\cite{HuaGli,GliSin,Musetal} are due to the inapplicability
of this theory, to the approximate nature of its
predictions in 3D, or to some missing 
physics in the starting equations. 

Here we report the results of 3D 
simulations of dendritic growth in a pure
undercooled melt. These simulations are made possible
by using a recently developed phase-field approach 
which renders fully quantitative
3D computations accessible for the first time
\cite{KarRap}. This approach also makes it possible 
to model the experimentally relevant low velocity limit 
where the solid-liquid interface 
can be assumed to relax instantaneously
to local thermodynamic equilibrium (i.e. where kinetic
effects at the interface are negligibly small).
Our numerical results are applied to critically test
the applicability of solvability theory in 3D, 
to make comparison with experiment, and to
characterize the full 3D dendrite 
tip morphology.

The equations of the 3D free-boundary problem 
are given by
\begin{eqnarray}
\partial_t T&=&D\nabla^2 T\\
Lv_n&=&c_pD\left(\left.\partial_n T\right|_S
-\left.\partial_n T\right|_L\right)\\
T_I-T_M&=&-\frac{T_M}{L}\,\sum_{i=1}^2\,
\left[\gamma({\bf n})+\frac{\partial^2\gamma({\bf n})}
{\partial \theta_i^2}\right]\frac{1}{R_i}\label{gt}
\end{eqnarray}
where $T$ is the temperature field,
$T_I$ is the interface temperature,
$T_M$ is the melting temperature,
$D$ is the diffusivity, $L$ is the latent heat of melting,
$c_p$ is the specific heat, $v_n$ is the
normal velocity of the interface, and
\begin{equation}
\gamma({\bf n})~=~\gamma_0(1-3\epsilon_4)\,
\left[1+\frac{4\epsilon_4}
{1-3\epsilon_4}\left(n_x^4+n_y^4+n_z^4\right)\right]\label{ani}
\end{equation}
is the surface energy for a crystal 
chosen to have a cubic symmetry.
In addition, $\theta_i$ are the angles between 
the normal ${\bf n}$ and the two local principal
directions on the boundary and $R_i$ are the 
principal radii of curvature. 
Growth is controlled by the 
undercooling $\Delta=(T_M-T_\infty)/(L/c_p)$
where $T_\infty$ is the initial melt
temperature.

We use three key ingredients 
to solve the above equations.
Firstly, we avoid the usual difficulties of tracking
a sharp boundary by using a
phase-field approach \cite{Lan,Kob,Wheetal,WanSek}.
In this approach, the two-phase system is
described by a phenomenological free-energy
\begin{equation}
{\cal F}=\int {\bf dr}\,[\,W^2({\bf n})|\nabla\psi|^2+f(\psi,u)]
\end{equation}
where $u\equiv (T-T_M)/(L/c_p)$ is
the dimensionless temperature field. We have used
here the form $f(\psi,u)=-\psi^2/2+\psi^4/4+\lambda\,u\,\psi\,
(1-2\psi^2/3+\psi^4/5)$ with minima at $\psi=-1$ and $\psi=+1$ 
that correspond to the liquid and solid phases, respectively.
The anisotropic surface energy defined by Eq. \ref{ani}
is recovered by choosing $W({\bf n})=W_0\gamma({\bf n})/\gamma_0$, with
${\bf n}\equiv\nabla \psi/|\nabla\psi|$, which is the direct
3D generalization of the way anisotropy has been previously included
in 2D \cite{Kob,Wheetal,WanSek,Macetal}.
The phase-field, $\psi$, varies between its two minima values
across an interfacial region of width $W_0$, thereby
distinguishing between phases without front-tracking.
Its dynamics, defined by 
\begin{equation}
\tau({\bf n})\,\frac{\partial\psi}{\partial t}=
-\frac{\delta {\cal
F}}{\delta \psi}, \label{p1}
\end{equation} 
is then coupled to that of the diffusion field
\begin{equation}
\frac{\partial u}{\partial t}
~=~D\,\nabla^2u+\frac{1}{2}\frac{\partial
\psi}{\partial t},\label{p2}
\end{equation}
in such a way that the equations 
for the two fields reduce to those of the original solidification
equations in the sharp-interface limit where 
$W_0$ is small compared to the
principal radii of curvature of the boundary. 

Secondly, and most importantly, 
we exploit the results of a recent analysis 
\cite{KarRap} to relate the phase-field 
model parameters to those of the
sharp-interface equations.
As demonstrated in 2D \cite{KarRap}, 
this analysis makes it possible:
(i) to perform computations with a smaller
capillary length, $d_0=\gamma_0T_Mc_p/L^2$, and
(ii) to choose $\lambda$ and the function
$\tau({\bf n})$ in such a way that we obtain 
a Gibbs-Thomson condition without interface kinetics 
in the sharp-interface limit \cite{KarRap}. That is we
recover exactly Eq. \ref{gt} without 
an additional velocity-dependent term in this limit.
Since the computation time scales
as $(d_0/W_0)^5$ in 3D, this first property dramatically enhances
the computational efficiency of the phase-field approach
and allows us to model dendritic growth quantitatively, 
as opposed to non-quantitatively in 3D as previously \cite{Kob}.
Furthermore, it permits us to model an undercooling range where the 
dendrite tip P\'eclet number $P\equiv \rho V/2D$
is small enough to compare $\sigma^*\equiv 2Dd_0/\rho^2V$
to its measured small undercooling 
values \cite{HuaGli,GliSin}. 

\begin{figure}
\def\epsfsize#1#2{0.4#1}
\newbox\boxtmp
\setbox\boxtmp=\hbox{\epsfbox{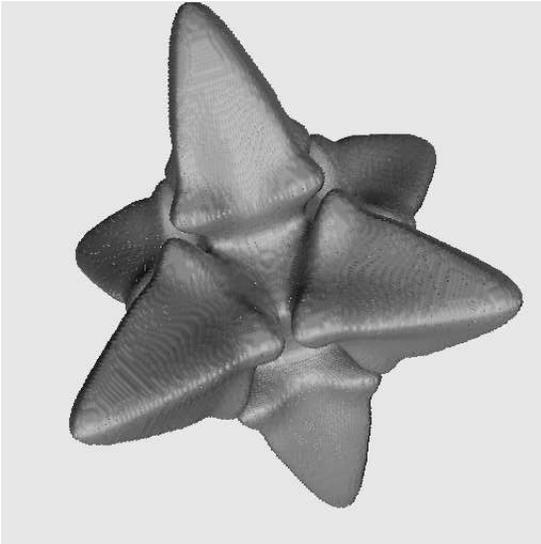}}
\rotr{\boxtmp}
\vspace{0.2cm}
\caption{
Results of 3D phase-field
simulation on a $300\times 300\times 300$ cubic lattice
for $\epsilon_4=0.047$ which shows dendrite tips
growing along the principal $<100>$ directions.
The simulations was performed in the
first octant $(x,y,z\ge 0)$ with a spherical nucleus 
centered at the origin as initial condition.
The solid-liquid boundary shown here corresponds to the $\psi=0$ 
surface reconstructed by reflection about 
the $x=y=z=0$ planes. The structure is seen from an angle
where all six $<100>$ directions are visible.} 
\label{nul1}
\end{figure}

Lastly, we are able 
to resolve numerically very small anisotropies 
by incorporating quantitatively the contribution of 
the grid anisotropy. For a given value of
$\epsilon_4$ used in $W({\bf n})$, we compute the equilibrium
shape produced by the phase-field model. The grid-corrected anisotropy
is then defined as that value of $\epsilon_4$ 
in $\gamma({\bf n})$ which produces 
a 3D equilibrium shape that matches 
exactly that of the phase-field model. A procedure
that will be described elsewhere was also 
developed to obtain a grid-corrected 
$\tau({\bf n})$. Numerical tests were performed in 2D 
to check that this procedure yields accurate values of 
$\sigma^*$ over the entire range
of $\epsilon_4$ investigated here in 3D.

Eqs. \ref{p1}-\ref{p2} 
were simulated explicitly with a
grid spacing ranging between
$0.6$ and $0.8$ with $W_0=1$, $D$ 
ranging between $0.5$ and $4$, 
a time step ranging between $0.01$ and $0.08$,
and a constant undercooling $\Delta=0.45$.
The anisotropy was varied from $\epsilon_4=0.0066$ 
to $\epsilon_4=0.047$ which spans most of the range
of experimental interest. Fig. 1 shows a typical
3D dendrite morphology resulting from the growth of
a small spherical seed. Quantitative results 
which pertain to the steady-state operating state
and morphology of the dendrite tip are presented 
in Table 1 and Figs. 2-3. 
These were obtained with long simulations that 
focused on the steady-state growth
of a single dendrite tip along the $z$-axis. 
Several runs were performed
to check that the values in Table 1 are 
independent of $d_0/W_0$ within 
an accuracy of 10\%. 

\begin{figure}
\def\epsfsize#1#2{0.4#1}
\newbox\boxtmp
\setbox\boxtmp=\hbox{\epsfbox{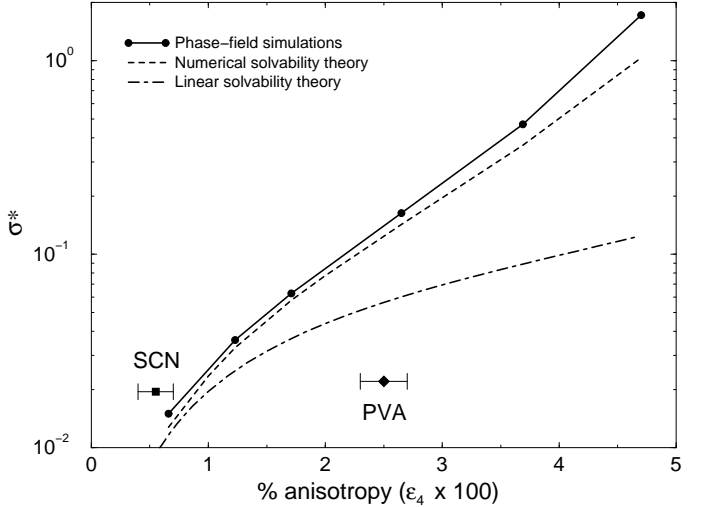}}
\rotr{\boxtmp}
\vspace{0.6cm}
\caption{
Plot of $\sigma^*$ vs $\epsilon_4$ for $\Delta=0.45$
showing the results of phase-field simulations
compared to the approximate predictions of
the numerical and linear solvability theories.
Also plotted are the small $\Delta$ experimental 
values of $\sigma^*$ for SCN [9]
and PVA [10] using the $\epsilon_4$ 
measurements of Ref. [11].}
\label{nul2}
\end{figure}

\begin{table}
\caption{Result of phase-field simulations
on a $200\times 200\times 400$ cubic
lattice compared to the results of the numerical and
linear solvability theories for $\Delta=0.45$. 
$A$ and $\alpha$ characterize 
the amplitude of the four-fold symmetry 
component of the tip morphology in simulations.
Typical runs took 60-140 CPU hours on
a DEC-ALPHA 3000-700 workstation and shorter times on
a CRAY-YMP and a CRAY-T3D.} 
\begin{tabular}{|c|cccc|cc|cc|} \hline
& \multicolumn{4}{c|}{Phase-Field Simulations} & 
\multicolumn{4}{c|}
{Solvability theory} \\ \cline{6-9}
 & & & & & \multicolumn{2}{c|}{Numerical} & 
\multicolumn{2}{c|}{Linear } \\ \hline
$\epsilon_4$ & $P$ & $\sigma^*$ & $\alpha$ & $A^{-1}$ 
& $P$ & $\sigma^*$ & $P_{Iv}$ & $\sigma^*$ \\ \hline
0.0066 & 0.426 & 0.015 & 1.78 & 13.0 & 0.418 & 0.0128 & 0.471 & 0.0116
\\
0.0123 & 0.360 & 0.036 & 1.73  & 11.3 & 0.367 & 0.0329 & 0.471 & 0.0250
\\
0.0171 & 0.312 & 0.063 & 1.68  & 10.1 & 0.324 & 0.0578 & 0.471 & 0.0367
\\
0.0265 & 0.236 & 0.16 & 1.62 & 7.8 & 0.247 & 0.142 & 0.471 & 0.0602 \\
0.0369 & 0.159 & 0.47 & 1.57 & 5.7  & 0.172 & 0.365 & 0.471 & 0.0890 \\
0.0470 & 0.093 & 1.72 & 1.54 & 4.0  & 0.109 & 1.037 & 0.471 & 0.1240 \\
\hline
\end{tabular}
\end{table}

In order to test solvability theory, we
have computed independently the values 
of $P$ and $\sigma^*$ predicted by the numerical
solution of the steady-state growth equations
using the standard axisymmetric approximation
where the surface energy and steady-state 
shape are assumed to be independent
of the polar angle $\phi$ in
the $x$-$y$ plane perpendicular to the growth axis
\cite{KesLev}. This is the same calculation performed
by Kessler and Levine in Ref. \cite{KesLev} and we
have checked that our boundary integral code 
reproduces their results within one percent.
For completeness, we also report the predictions
of the linear solvability theory
of Barbieri and Langer \cite{BarLan} which uses a shape
linearized around the Ivantsov paraboloid of revolution
\cite{Iva} and the same axisymmetric approximation.

As shown in Table 1, the numerical
solvability theory yields $\sigma^*$ values
which are systematically lower than the
phase-field results, but still reasonably close
for small anisotropy. The predicted values
of $\sigma^*$, however, start to 
become significantly inaccurate
for anisotropies greater than about 3\%, although
predictions of the  P\'eclet number remain relatively accurate.
Most likely, this does not indicate a breakdown of
solvability theory, but of the
axisymmetric approximation. This conclusion 
is supported by the fact that
the four-fold deviation from a shape of revolution 
increases in magnitude with anisotropy (as described below)
and, hence, can affect the selection. 
It is also supported by the fact that we do
not observe any sidebranching \cite{LanHou,HuaGli,Douetal} 
without adding noise to the phase-field equations.
Hence our simulations rule out the possibility of
a dynamical attractor other than a steady-state
needle crystal. The linear theory is seen to breakdown at
much smaller anisotropy. This is because
it assumes that the steady-state shape remains close
to the Ivantsov paraboloid of revolution. 
Table 1 shows that the actual P\'eclet number 
already starts deviating significantly from its Ivantsov value,
$P_{Iv}$ \cite{Iva}, at small anisotropy.

The steady-state morphology of the 
dendrite tip was analyzed
using the Fourier decomposition 
\begin{equation}
r^2(\phi,z)=\sum_n\,A_n(z)\cos 4n\phi
\end{equation}
where $r$ is the radial distance 
from the $z$-axis. Both $r$ and $z$ are
measured in units of $\rho$ with the tip at $z=0$.
This decomposition has the advantage that it is general 
and does not presupposes a particular analytical form to 
fit the tip shape. The function
$A_0(z)$ is the axisymmetric contribution to
this shape. It approaches a paraboloid
of revolution, $A_0(z)=-2\,z$, for small $|z|$
and departs from this shape with increasing $|z|$
as shown in Fig. 3, this departure being more
pronounced at larger anisotropy as one would expect.
More important is the non-axisymmetric departure from
this shape of revolution which is contained in the
higher modes, $A_n(z)$ for $n\ge 1$. 
The amplitude of the
first four-fold symmetry mode turns out to be much larger than all
the other modes (Fig. 3) and to be well-described by the 
power law $A_1(z)=A\,|z|^\alpha$, which appears
as a remarkably straight line 
on a log-log plot of $A_1(z)$ vs $|z|$ up to a distance 
of a few $\rho$ behind the tip. 
Furthermore, the values of $A$ and $\alpha$ 
in Table 1 clearly show that the amplitude of the four-fold
symmetry mode is sensitively dependent on
anisotropy which is a qualitatively novel aspect 
of our results. In contrast, the
linear solvability calculation of Ben Amar 
and Brener \cite{BenBre} predicts that for 
small $\epsilon_4$ the tip morphology should be 
independent of anisotropy, with $A^{-1}=11$ and $\alpha=2$
for small $|z|$. The values 
of $A$ and $\alpha$ listed in Table 1 
indicate that this prediction is most likely only valid
for values of $\epsilon_4$ which lie outside the
range of experimental interest. This is also consistent with
the fact that their calculation is only valid in the limit where
$\sigma^*\sim \epsilon_4^{7/4}$, and that this $7/4$ power law
scaling is not yet attained at the smallest computed
anisotropy in Fig. 2.

For $\epsilon_4=0.0066$, which lies
inside the range of uncertainty of the
measured value $\epsilon_4=0.0055\pm 0.0015$ for SCN \cite{Musetal},
our simulations yield a value of $\sigma^*\approx 0.015$ 
with $P=0.426$. This value is
only 20 \% smaller than the value $\sigma^*=0.0192$ measured
by Huang and Glicksman \cite{HuaGli} for this material.
Most of this discrepancy can be accounted for
by the finite P\'eclet number correction which can be estimated,
using the numerical solvability code, to decrease $\sigma^*$
by about 15\% from its zero P\'eclet number limiting
value. Therefore we obtain a reasonably
good agreement with experiment for 
SCN within the existing uncertainty in the
measured value of anisotropy.
Pivalic acid \cite{GliSin} (PVA), however, remains 
problematic as seen in Fig. 2 and a closer 
examination of kinetic effects for this 
material could potentially resolve this large discrepancy.
Finally, Maurer {\it et al.} \cite{Mauetal} have found
that the tip morphology of NH$_4$Br dendrites 
is indeed well-described by a
single $\cos\,4\phi$ mode, while LaCombe {\it et al.}
have found that for SCN dendrites more modes seem
necessary to fit the tip morphology \cite{LaCetal}.
The origin of this difference, which is potentially
due to noise amplification, also remains to be understood.

In conclusion, we have 
demonstrated that quantitative 
modeling of 3D dendritic growth is possible
using a recently developed 
phase-field methodology \cite{KarRap}.
Our results are important in 
that they leave little doubt about the 
conceptual validity of solvability theory in 3D and
the importance of crystalline anisotropy,
and yield results which are consistent with experiment, at least
for SCN where kinetic effects are believed to be small.
At present, the applicability of solvability theory $-$
in terms of making accurate predictions $-$ remains
mainly limited by the axisymmetric approximation.
Our results have shown that the three-dimensional morphology
of the dendrite tip is dominated by a 
four-fold component whose amplitude
depends sensitively on anisotropy. This prediction should
be testable experimentally. A host of other microstructural
pattern formation issues, such as the determination of the
steady-state shape further behind the tip and sidebranching,
are now amenable to quantitative study 
using the present computational approach.

This research was supported by a DOE
grant and benefited from supercomputer time allocation at NERSC.
We thank Stuart Levy of the Geometry Center at the University of
Minnesota for his help in visualizing our results and Martine
Ben Amar for useful exchanges.

\begin{figure}
\def\epsfsize#1#2{0.45#1}
\newbox\boxtmp
\setbox\boxtmp=\hbox{\epsfbox{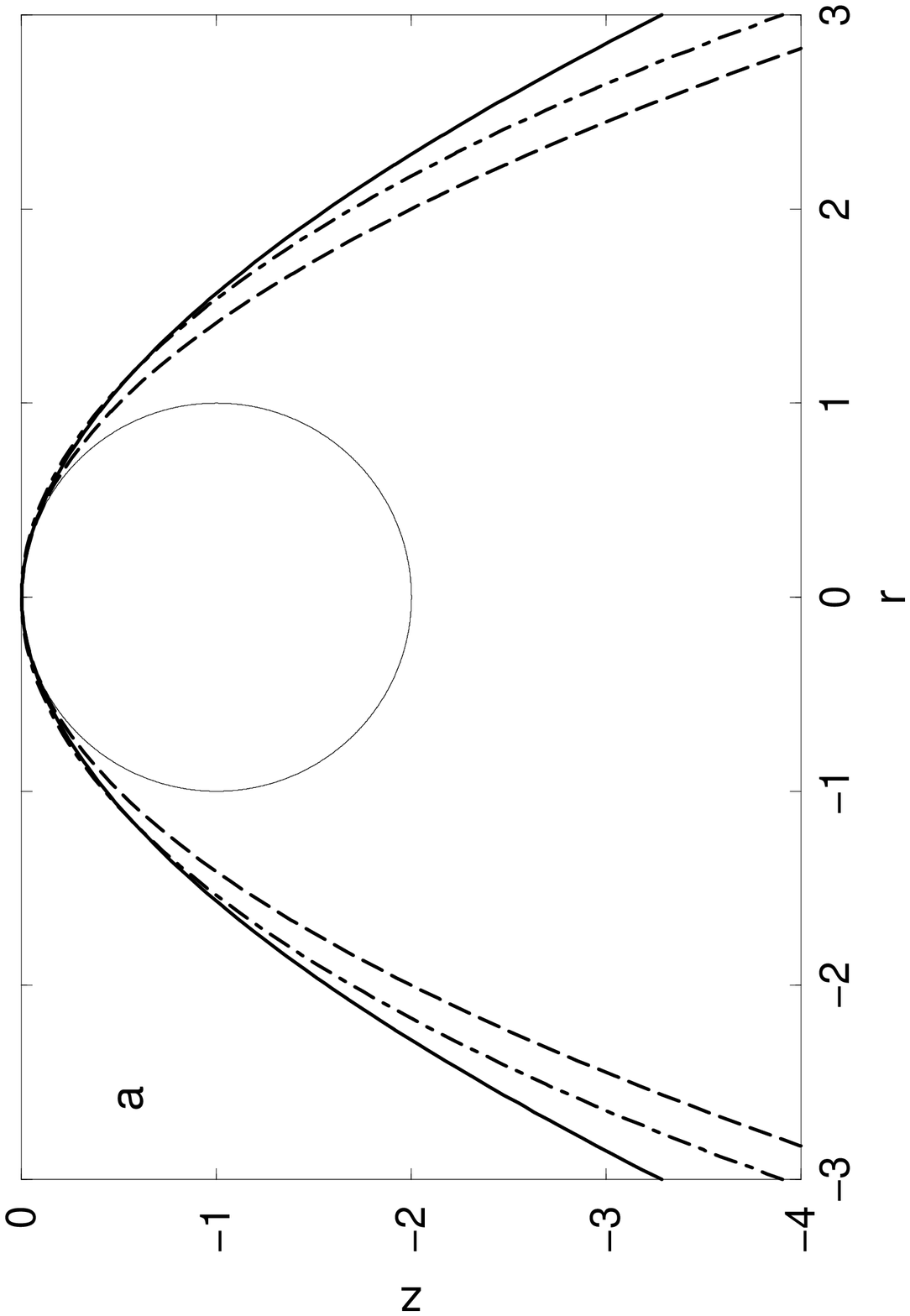}}
\rotr{\boxtmp}
\label{nul3a}
\end{figure}

\begin{figure}
\def\epsfsize#1#2{0.5#1}
\newbox\boxtmp
\setbox\boxtmp=\hbox{\epsfbox{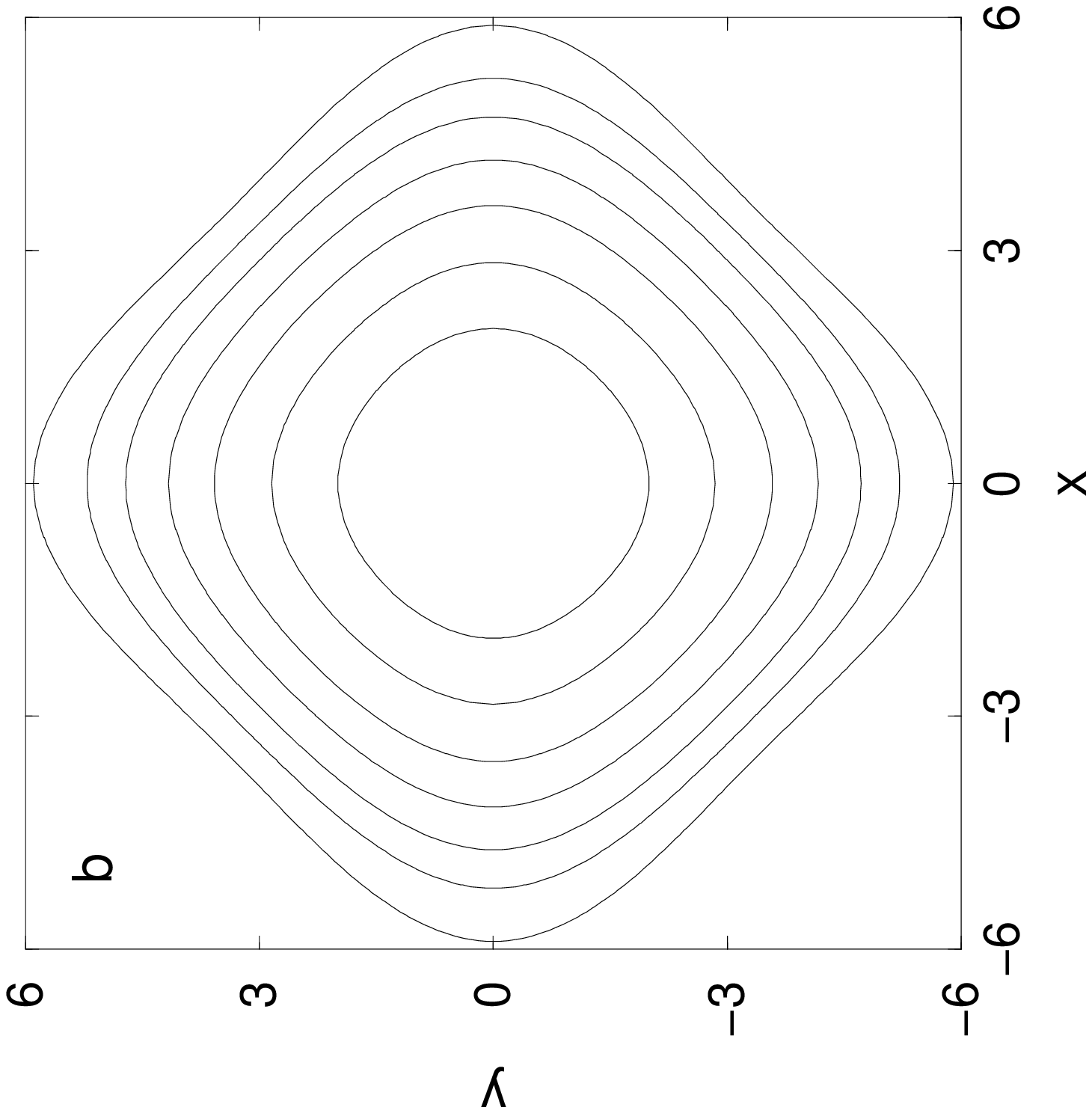}}
\rotr{\boxtmp}
\vspace{0.6cm}
\caption{
Steady-state tip morphology 
for $\epsilon_4=0.0123$ shown in: (a) the $\phi=0^o$ (solid line)
and $\phi=45^o$ (dash-dotted line) planes, and (b)
(100) planes equally spaced along $z$ by one tip radius.
Length is measured in units of $\rho$.
The Ivantsov parabola $z=-r^2/2$ (dash-line)
and a circle of unit radius are 
superimposed in (a). The Fourier amplitudes 
at two tip radii behind the tip in (b) are:
$A_1(-2)=0.29$, $A_2(-2)=0.020$, and $A_3(-2)=0.0022$,
illustrating that the tip morphology is dominated 
by the four-fold symmetry mode.}
\label{nul3b}
\end{figure}

\end{document}